\documentclass[conference]{IEEEtran}
\IEEEoverridecommandlockouts

\usepackage{cite}
\usepackage{amsmath,amssymb,amsfonts}
\usepackage{graphicx}
\usepackage{textcomp}
\usepackage{xcolor}
\usepackage[hidelinks]{hyperref}
\usepackage{orcidlink}
\usepackage{svg}
\usepackage{booktabs}

\def\BibTeX{{\rm B\kern-.05em{\sc i\kern-.025em b}\kern-.08em
    T\kern-.1667em\lower.7ex\hbox{E}\kern-.125emX}}

\begin{document}

\title{CAMTA: A Reconfigurable Multi-Region Activation Unit for Nonlinear Function Approximation}

\author{
   \IEEEauthorblockN{
     Carlos~Soto-Porras\orcidlink{0000-0002-1569-2199},
     Jose~Fonseca-Cruz\orcidlink{0000-0002-9529-3576},
     Pablo~Ram\'irez-Morera\orcidlink{0009-0004-2291-8880},
     Erick~Obreg\'on-Fonseca\orcidlink{0009-0007-1046-0894},\\
     Luis~G.~Le\'on-Vega\orcidlink{0000-0002-3263-7853},
     Jorge~Castro-God\'{i}nez\orcidlink{0000-0003-4808-4904}
   }
   \vspace{3pt}
   \IEEEauthorblockA{School of Electronics Engineering, Costa Rica Institute of Technology, Costa Rica}
}

\maketitle

\begin{abstract}
Nonlinear activation functions are widely used in machine learning workloads, but their direct hardware implementation is often costly, function-specific, or difficult to reuse across different models. This work introduces CAMTA, a 16-bit reconfigurable multi-region activation unit for nonlinear function approximation in FPGA and ASIC accelerators. CAMTA combines independent region thresholds, per-region polynomial degrees, coefficient sets, and execution modes over a shared Horner-based datapath. Unlike conventional polynomial or piecewise approximation units that mainly reconfigure coefficients or segment selection, CAMTA also reconfigures the computational behavior of each region through HORNER, CONST, ZERO, and IDENTITY modes, enabling the same hardware to support functions with different symmetry and tail behavior without resynthesis. FPGA validation on an AMD Alveo platform shows RMSE as low as \(3.60\times10^{-6}\) for CAMTA-assisted Softmax, outperforming the CORDIC-based Softmax baseline considered in this work by nearly one order of magnitude. FPGA HLS synthesis reports 3 DSPs, 802 FFs, 1756 LUTs, and an 11-cycle datapath latency. ASIC synthesis in TSMC 65~nm at 250~MHz reports \(6632.40~\mu\mathrm{m}^2\) total cell area and \(1.3634~\mathrm{mW}\) total power. Compared with a same-node, function-specific PLAC implementation, CAMTA incurs \(2.20\times\) area and \(1.75\times\) power overhead, in exchange for runtime configurability and reuse across multiple nonlinear functions.
\end{abstract}


\section{Introduction}

Nonlinear functions, such as GeLU, tanh, sigmoid, Swish, and Softmax, are central to modern neural network training and inference. Still, they remain expensive to implement directly in hardware due to their transcendental or high-curvature behavior \cite{KimUnifiedApprox, LiSoftmaxGELUViT, Carrillo2026SoftmaxCORDIC}. Specialized accelerators can achieve high accuracy and low latency for a target function, but their datapaths are usually tied to a specific operation and must be redesigned or resynthesized when the activation changes.

Approximate computing provides an alternative when small numerical deviations are acceptable~\cite{leon2022dcas,leon2023tecs}. Fixed-point piecewise linear, polynomial, LUT-based, and CORDIC-based methods have all been used to reduce nonlinear functions to hardware-friendly arithmetic \cite{DongPLAC, Dalloo2026OtLUT, volder, Walther}. However, the main design challenge is not only accuracy, but also reusability and flexibility: inference accelerators may need to support several activations across different models without duplicating dedicated hardware for each one. Also,  fabricating a chip has high investment costs, and it is also time-consuming. Even when FPGAs have advantages for fast implementation, they are less energy-efficient and slower than equivalent ASIC implementations \cite{kachris2025survey}.

Hence, this work proposes CAMTA, a reconfigurable multi-region activation unit that approximates several nonlinear functions using the same scalar datapath. Unlike a function-specific unit, CAMTA exposes runtime-configurable thresholds, coefficients, polynomial degrees, and region modes. 

The main contributions of this work are:

\begin{itemize}
    \item a scalar 16-bit reconfigurable activation unit which supports multiple nonlinear functions through a shared datapath, without the need for function-specific specialized hardware;
    \item a multi-level reconfiguration scheme based on independent region thresholds, per-region polynomial degrees, coefficient sets, and execution modes, enabling the unit to adapt to different non-linear behaviors;
    \item a mode-controlled regional bypass that extends conventional polynomial evaluation by allowing each region to operate as HORNER, CONST, ZERO, or IDENTITY, reducing unnecessary Horner evaluation in asymptotic or low-curvature regions;
\end{itemize}

These contributions showcase the main advantage of CAMTA over other contributions, as there is no need for re-synthesis and silicon re-spin depending on the required non-linear function; hence, giving the flexibility and also the robustness of keeping track with future machine learning algorithms when newer and improved activation functions arise.

\section{Related Work}

Hardware nonlinear-function accelerators can be broadly grouped into specialized evaluators and approximation-based configurable units. Specialized Softmax or exponential accelerators, including CORDIC-based and split-based architectures, provide strong latency and accuracy for specific functions but do not directly generalize to arbitrary activation functions \cite{BasicSplit, cordic, Carrillo2026SoftmaxCORDIC}. CORDIC remains attractive for elementary functions due to its regular iterative structure \cite{volder, Walther}, but it is not designed as a polynomial activation unit configurable across several smooth nonlinearities.

Piecewise linear approximation methods, such as PLAC, use segmentation and quantization strategies to approximate nonlinear unary functions with controlled error \cite{DongPLAC}. These methods are effective for many monotone functions and can reduce redesign effort, but PWL approximations may require several segments to capture high-curvature regions. Piecewise polynomial approaches improve local fitting capability at the cost of additional arithmetic. Hardware activation-function cores based on Horner evaluation and configurable coefficients have been proposed to support multiple nonlinear functions on a shared datapath \cite{GonzalezAFC, KimUnifiedApprox}. CAMTA follows this configurable-core direction, but introduces independent region boundaries and an explicit region mode to bypass the polynomial datapath when a tail can be represented as a constant, zero, or identity mapping, thereby supporting truncation and ReLU activation functions for self-contained computation.

\section{CAMTA Architecture}

\begin{figure}[t]
    \centering
    \includegraphics[width=0.98\linewidth]{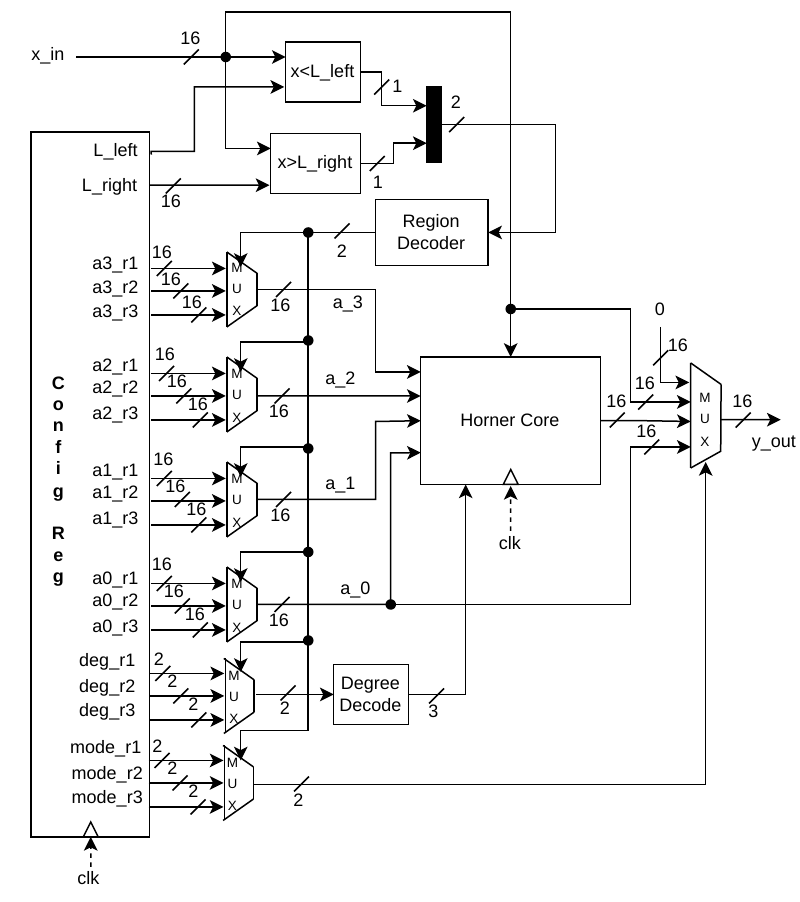}
    \vspace{-5mm}
    \caption{CAMTA scalar activation unit with independent thresholds, coefficient and degree selection, and mode-controlled output bypass.}
    \vspace{-5mm}
    \label{fig:camta_arch}
\end{figure}

CAMTA is a scalar configurable activation unit with 16-bit Q6.10 fixed-point input and output, and a wider internal accumulator in the Horner chain to reduce intermediate quantization effects. Fig.~\ref{fig:camta_arch} shows that the input \(x_{in}\) is compared against two independent thresholds, \(L_{left}\) and \(L_{right}\), defining three execution regions:

\begin{equation}
    r_0: x < L_{left}, \quad
    r_1: L_{left} \leq x \leq L_{right}, \quad
    r_2: x > L_{right}.
\end{equation}

The resulting region selector drives the multiplexers that choose the active coefficient set, polynomial degree, and execution mode.

For each region \(r_i\), CAMTA receives four coefficients \(a_{3,i}\), \(a_{2,i}\), \(a_{1,i}\), \(a_{0,i}\), a polynomial degree \(d_i \in \{0,1,2,3\}\), and a mode \(m_i\). In HORNER mode, the selected region is evaluated as:

\begin{equation}
P_i(x)=a_{0,i}+a_{1,i}x+a_{2,i}x^2+a_{3,i}x^3,
\end{equation}

implemented through Horner's rule:

\begin{equation}
P_i(x)=(((a_{3,i}x+a_{2,i})x+a_{1,i})x+a_{0,i}).
\end{equation}

The degree decoder disables unused polynomial stages, allowing the same core to implement constant, linear, quadratic, or cubic approximations.

\begin{figure}[t]
    \centering
    \includegraphics[width=\linewidth]{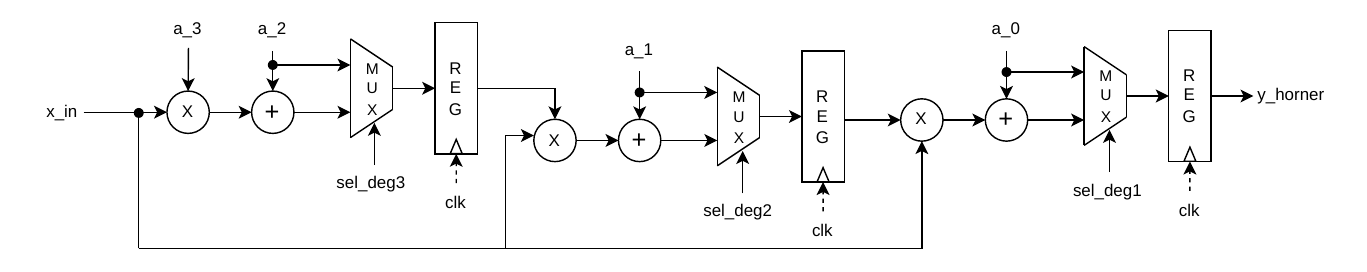}
    \caption{Degree-controlled Horner core. The output is \(y_{horner}\), which is later selected or bypassed by the CAMTA output mux.}
    \label{fig:horner_core}
\end{figure}

The region mode controls the final output mux. In HORNER mode, the unit returns \(y_{horner}\); in CONST mode, it returns the selected \(a_0\) coefficient; in ZERO mode, it returns zero; and in IDENTITY mode, it returns \(x_{in}\). This mode-controlled bypass extends conventional polynomial activation units by avoiding arithmetic evaluation in regions where the function can be represented by simpler behavior, such as constant, near-zero, or identity-like tails.

A key architectural feature is that CAMTA not only reconfigures coefficients or segment boundaries. It also reconfigures the computational behavior of each region. This allows a single shared Horner-based datapath to approximate functions with different symmetry and tail characteristics without instantiating dedicated hardware per activation. For example, tanh and sigmoid can use constant tails, while GeLU and Swish can use zero or identity-like tails. Thus, CAMTA combines region selection, degree selection, coefficient selection, and output-mode selection into a single reusable scalar unit.

\section{Results and Analysis}

\subsection{FPGA and ASIC Unit Synthesis}

The CAMTA scalar unit\footnote{Available: \url{https://github.com/ECASLab/camta-non-linear-unit}} was first synthesized as an FPGA-oriented HLS block for the AMD Alveo U55C. For the ASIC evaluation, the generated RTL was isolated to retain only the arithmetic and control core, excluding HLS-generated interface logic and wrapper components, and then synthesized using a TSMC 65~nm standard-cell library. The FPGA result characterizes the unit as a reusable activation datapath, while the ASIC result evaluates its feasibility as a compact standard-cell macro. The HLS synthesis reports a 4.00~ns target clock, a 2.465~ns estimated clock, and an 11-cycle datapath latency. Table~\ref{tab:fpga_synth} compares the FPGA resource usage of CAMTA against a reported CORDIC-based exponential core \cite{Carrillo2026SoftmaxCORDIC}. This comparison should be interpreted carefully: the CORDIC result applies only to an exponential evaluator, whereas CAMTA supports multiple nonlinear functions via runtime-configurable thresholds, coefficients, degrees, and region modes. Even with this broader functionality, CAMTA uses fewer LUTs than the CORDIC exponential core, at the cost of one additional DSP and a slightly higher FF count.

\begin{table}[t]
\centering
\caption{FPGA resource comparison with a CORDIC exponential core.}
\label{tab:fpga_synth}
\scriptsize
\renewcommand{\arraystretch}{1.10}
\setlength{\tabcolsep}{5pt}
\begin{tabular}{@{}lcc@{}}
\toprule
\textbf{Metric} & \textbf{CAMTA} & \textbf{CORDIC exp. core} \\
\midrule
Target clock & 4.00 ns & -- \\
Estimated clock & 2.465 ns & -- \\
Latency & 11 cycles & -- \\
BRAM & 0 & -- \\
DSP & 3 & 2 \\
FF & 802 & 757 \\
LUT & 1756 & 2676 \\
\bottomrule
\end{tabular}
\end{table}

The same unit was also synthesized using a TSMC 65~nm standard-cell flow at a 4.0~ns clock period, equivalent to 250~MHz. Table~\ref{tab:asic_synth} reports the ASIC implementation result. The design achieved a total cell area of 6632.40~$\mu m^2$, with 4555.60~$\mu m^2$ combinational area and 2076.80~$\mu m^2$ noncombinational area.
The power report shows 1.1743~mW of dynamic power, 189.16~\(\mu\)W of leakage power, and 1.3634~mW of total power.

\begin{table}[t]
\centering
\caption{TSMC 65~nm synthesis result for CAMTA.}
\label{tab:asic_synth}
\scriptsize
\renewcommand{\arraystretch}{1.10}
\setlength{\tabcolsep}{5pt}
\begin{tabular}{@{}lc@{}}
\toprule
\textbf{Metric} & \textbf{Value} \\
\midrule
Clock period & 4.0 ns \\
Frequency & 250 MHz \\
Critical path & 2.62 ns \\
\midrule
Total cell area & \(6632.40~\mu m^2\) \\
Combinational area & \(4555.60~\mu m^2\) \\
Noncombinational area & \(2076.80~\mu m^2\) \\
Total cells & 1582 \\
Combinational cells & 1343 \\
Sequential cells & 236 \\
\midrule
Horner core area & \(4753.60~\mu m^2\) \\
Region-selection area & \(164.00~\mu m^2\) \\
Mode-bypass area & \(83.20~\mu m^2\) \\
\midrule
Dynamic power & 1.1743 mW \\
Leakage power & 189.16 \(\mu\)W \\
Total power & 1.3634 mW \\
\bottomrule
\end{tabular}
\end{table}

The hierarchical area report shows that the Horner core dominates the implementation area, occupying approximately 71.7\% of the total cell area. In contrast, the region-selection and mode-bypass logic occupy only 2.5\% and 1.3\%, respectively. This indicates that the added configurability and output bypass introduce a small area overhead compared with the arithmetic datapath.

Table~\ref{tab:asic_comparison} compares CAMTA against the closest same-node PLAC result reported in TSMC 65~nm~\cite{DongPLAC}. The comparison must be interpreted with the corresponding architectural context: PLAC implements a function-specific piecewise-linear approximation circuit for \(\log_2(1+x)\), while CAMTA is a reconfigurable scalar macro intended to support multiple nonlinear functions through runtime-configurable thresholds, coefficients, polynomial degrees, and output modes. CAMTA requires approximately \(2.20\times\) the area and \(1.75\times\) the power of the reported PLAC 65~nm implementation, but this overhead is exchanged for flexibility and hardware reuse across several activation functions.

\begin{table}[t]
\centering
\caption{ASIC comparison with a same-node PLAC implementation.}
\label{tab:asic_comparison}
\scriptsize
\renewcommand{\arraystretch}{1.10}
\setlength{\tabcolsep}{4pt}
\begin{tabular}{@{}lcc@{}}
\toprule
\textbf{Metric} & \textbf{CAMTA} & \textbf{PLAC} \\
\midrule
Functionality & Reconfigurable & Function-specific \\
Reference function & Multiple AFs & \(\log_2(1+x)\) \\
Cell area & \(6632.40~\mu m^2\) & \(3019~\mu m^2\) \\
Total power & 1.3634 mW & 0.7812 mW \\
Critical path / delay & 2.62 ns & 1.48 ns \\
\bottomrule
\end{tabular}
\end{table}

\subsection{FPGA Runtime and Accuracy Validation}

The configured unit was validated through an AMD Alveo U55C FPGA wrapper using uniformly sampled inputs and floating-point software references. Table~\ref{tab:camta_runtime_results} reports representative runtime and numerical results for the best obtained configuration of each evaluated function using 10000 samples. The reported runtime corresponds to the scalar FPGA wrapper execution and should not be interpreted as the intrinsic latency of the CAMTA unit, since the wrapper includes memory-mapped execution and launch overhead. The intrinsic datapath latency is reported separately in Table~\ref{tab:fpga_synth}.

\begin{table}[t]
\centering
\caption{Representative runtime and numerical results for CAMTA with \(N=10000\).}
\label{tab:camta_runtime_results}
\scriptsize
\renewcommand{\arraystretch}{1.10}
\setlength{\tabcolsep}{3.2pt}
\begin{tabular}{@{}lccccc@{}}
\toprule
\textbf{Function} & \textbf{Kernel} & \textbf{MS/s} & \textbf{RMSE} & \textbf{MaxAE} \\
 & & \textbf{[\(\mu\)s]} & & & \\
\midrule
GeLU & 183.59 & 54.47 & 0.00152 & 0.00501 \\
tanh & 185.74 & 53.84 & 0.00162 & 0.00582 \\
sigmoid & 185.89 & 53.79 & 0.00207 & 0.00803 \\
Swish & 185.42 & 53.93 & 0.00389 & 0.01344 \\
Softmax$^\dagger$ & 185.24 & 53.98 & \(3.60{\times}10^{-6}\) & \(1.08{\times}10^{-5}\) \\
\bottomrule
\end{tabular}

\vspace{2pt}
\footnotesize{\(^{\dagger}\)CAMTA-assisted Softmax: CAMTA approximates the exponential stage over \([-8,0]\), while summation and normalization are performed on the host.}
\end{table}

The best configurations exploit the proposed region modes according to the behavior of each activation function. GeLU and Swish use residual symmetry with identity-like tails, tanh uses odd symmetry with a constant tail, and sigmoid uses complement symmetry with a constant tail. For the Softmax experiment, CAMTA approximates the exponential stage over inputs shifted to the range \([-8,0]\), following the common formulation in which the maximum input value is subtracted before exponentiation. The summation and normalization stages are performed on the host; therefore, the result is reported as CAMTA-assisted Softmax rather than as a standalone hardware Softmax unit. Table~\ref{tab:camta_configs} summarizes the selected configurations.

\begin{table}[t]
\centering
\caption{Selected CAMTA configurations used in the \(N=10000\) experiments.}
\label{tab:camta_configs}
\scriptsize
\renewcommand{\arraystretch}{1.10}
\setlength{\tabcolsep}{2.2pt}
\begin{tabular}{@{}lcccccc@{}}
\toprule
\textbf{Function} & \textbf{Range} & \(\mathbf{L_l}\) & \(\mathbf{L_r}\) & \textbf{Sym.} & \textbf{Modes} & \textbf{Deg.} \\
\midrule
GeLU & \([-8,8]\) & 1.50 & 3.00 & residual & H/H/I & 3/3/1 \\
tanh & \([-4,4]\) & 1.25 & 3.50 & odd & H/H/C & 3/3/0 \\
sigmoid & \([-8,8]\) & 2.50 & 4.50 & comp. & H/H/C & 3/3/0 \\
Swish & \([-8,8]\) & 2.00 & 6.50 & residual & H/H/I & 3/3/1 \\
Softmax$^\dagger$ & \([-8,0]\) & -4.00 & -1.00 & none & C/H/H & 0/3/3 \\
\bottomrule
\end{tabular}

\vspace{2pt}
\footnotesize{H: HORNER, C: CONST, I: IDENTITY. \(L_l\) and \(L_r\) denote the left and right thresholds. For Softmax, the configuration corresponds to the exponential approximation stage.}
\end{table}

These results show that the same CAMTA datapath can approximate bounded, non-bounded, and exponential-based nonlinear operations by modifying only thresholds, coefficients, polynomial degrees, and region modes. This keeps the proposed unit independent from a specific memory-bus width while allowing system designers to scale throughput according to the target accelerator, preserving flexibility and reusability while maintaining bounded approximation error across multiple nonlinear functions.

\subsection{Comparison with Related Approximators}

Table~\ref{tab:comparison} compares CAMTA with related nonlinear-function approximators. For unary activations, CAMTA is compared using MaxAE, as PLAC and AFC/PPA-ED report maximum-error metrics~\cite{DongPLAC, GonzalezAFC}. For Softmax, RMSE is used following the CORDIC-based Softmax study~\cite{Carrillo2026SoftmaxCORDIC}. The Softmax result is CAMTA-assisted because CAMTA approximates the exponential stage, while reduction and normalization are performed on the host.

\begin{table}[t]
\centering
\caption{Comparison with related nonlinear-function approximators.}
\label{tab:comparison}
\scriptsize
\renewcommand{\arraystretch}{1.08}
\setlength{\tabcolsep}{2.1pt}
\begin{tabular}{@{}lcccccc@{}}
\toprule
\textbf{Op.} & \textbf{Metric} & \textbf{CAMTA} & \textbf{PLAC} & \textbf{AFC} & \textbf{CORDIC} & \textbf{Taylor / Lin. Intp.} \\
\midrule
\(\tanh\) &
MaxAE \((10^{-3})\) &
5.824 & 5.55 & 5.90 & -- & -- \\

\(\sigma\) &
MaxAE \((10^{-3})\) &
8.035 & 5.65 & 2.10 & -- & -- \\

Swish &
MaxAE \((10^{-3})\) &
13.44 & -- & 7.90 & -- & -- \\

Softmax\(^{\dagger}\) &
RMSE \((10^{-6})\) &
3.60 & -- & -- & \(\sim30\) & \(>10^{4}\) / \(\sim5000\) \\
\bottomrule
\end{tabular}

\vspace{2pt}
\footnotesize{\(^{\dagger}\)CAMTA-assisted Softmax. CAMTA approximates the exponential stage; reduction and normalization are not implemented inside the CAMTA unit. Softmax baselines are approximated from the RMSE trends in~\cite{Carrillo2026SoftmaxCORDIC}: CORDIC \(\sim3\times10^{-5}\), Taylor \(>10^{-2}\), and Linear Interpolation \(\sim5\times10^{-3}\).}
\end{table}

CAMTA remains close to PLAC and AFC for \(\tanh(x)\), while AFC/PPA-ED reports lower worst-case error for sigmoid and Swish/SILU. In the Softmax case, the Horner-based exponential approximation achieves lower RMSE than the CORDIC, Taylor-based, and interpolation-based approximators considered in~\cite{Carrillo2026SoftmaxCORDIC}. These results highlight the main trade-off: CAMTA is not always the minimum-error evaluator for a single function, but it provides a reusable datapath that can be reconfigured across different nonlinear behaviors through thresholds, coefficients, polynomial degrees, and region modes. In actual AI models, the error might be absorbed by post-training fine-tuning.

\section{Conclusion}

This work presents CAMTA, a reconfigurable multi-region activation unit for nonlinear function approximation. CAMTA combines independent thresholds, per-region coefficients, configurable polynomial degrees, and mode-controlled output behavior over a shared Horner-based datapath. Horner evaluation enables up to cubic polynomial approximation with a compact multiply-accumulate structure, while bypass modes avoid unnecessary polynomial evaluation in CONST, ZERO, or IDENTITY regions. FPGA synthesis reports 3 DSPs, 802 FFs, 1756 LUTs, and no inferred BRAM, while ASIC synthesis in TSMC 65~nm shows operation at 250~MHz with \(6632.40~\mu\mathrm{m}^2\) cell area and \(1.3634~\mathrm{mW}\) total power. Runtime validation with \(N=10000\) samples shows competitive accuracy across GeLU, tanh, sigmoid, Swish, and CAMTA-assisted Softmax. Compared with function-specific approximators such as PLAC, CAMTA introduces area and power overhead, which is exchanged for runtime configurability, multi-function reuse, and support for different nonlinear tail behaviors without hardware resynthesis. The approximation results remain competitive with function-specific alternatives, and the impact of the approximations will be left to future work.

\section{Acknowledgements}

This work was supported by the Costa Rica Institute of Technology under the research project 1360058 (Generaci\'on Autom\'atica de Hardware para Aplicaciones de Aprendizaje Autom\'atico basadas en FPGA). The results of this work were partially supported by AMD through the Heterogeneous Accelerated Compute Clusters (HACC) program.
\bibliographystyle{IEEEtran}
\bibliography{references}

@INPROCEEDINGS{cordic,
  author={Cao, Yongxiang and others},
  booktitle={2020 IEEE International Conference on Artificial Intelligence and Information Systems (ICAIIS)}, 
  title={{Cordic-based Softmax Acceleration Method of Convolution Neural Network on FPGA}}, 
  year={2020},
  volume={},
  number={},
  pages={66-70},
  doi={10.1109/ICAIIS49377.2020.9194894}}

@ARTICLE{volder,
  author={Volder, Jack E.},
  journal={IRE Transactions on Electronic Computers}, 
  title={The CORDIC Trigonometric Computing Technique}, 
  year={1959},
  volume={EC-8},
  number={3},
  pages={330-334},
  doi={10.1109/TEC.1959.5222693}}

@inproceedings{Walther,
author = {Walther, J. S.},
title = {A unified algorithm for elementary functions},
year = {1971},
isbn = {9781450379076},
doi = {10.1145/1478786.1478840},
booktitle = {Proceedings of the May 18-20, 1971, Spring Joint Computer Conference},
pages = {379–385},
numpages = {7},
location = {Atlantic City, New Jersey},
}

@article{Dalloo2026OtLUT,
  author  = {Dalloo, Ayad M. and Waiss, Sulaf and Humaidi, Amjad J. and Al Mhdawi, Ammar K. and Noaman, Noaman M. and Ahmed, Husham M. and Mahdi Al-Obaidi, Abdulkareem Sh.},
  title   = {An Optimized Twofold LUT architecture for fast function approximation},
  journal = {Franklin Open},
  volume  = {15},
  pages   = {100489},
  year    = {2026},
  doi     = {10.1016/j.fraope.2026.100489}
}

@article{LiSoftmaxGELUViT,
  author  = {Li, Tianyang and Zhang, Fan and Xie, Guangwei and Fan, Xitian and Gao, Yanzhao and Sun, Mingqian},
  title   = {{A high speed reconfigurable architecture for softmax and GELU in vision transformer}},
  journal = {Electronics Letters},
  volume = {59},
  number = {5},
  year    = {2023},
  doi     = {10.1049/ell2.12751}
}

@article{GonzalezAFC,
  author  = {Gonz{\'a}lez-D{\'i}az Conti, Griselda and V{\'a}zquez-Castillo, Javier and Longoria-Gandara, Omar and Castillo-Atoche, Alejandro and Carrasco-Alvarez, Roberto and Espinoza-Ruiz, Adolfo and Ruiz-Ibarra, Erica},
  title   = {Hardware-Based Activation Function-Core for Neural Network Implementations},
  journal = {Electronics},
  volume  = {11},
  number  = {1},
  pages   = {14},
  year    = {2022},
  doi     = {10.3390/electronics11010014}
}

@article{KimUnifiedApprox,
  author  = {Kim, Jeongmin and Choi, Kangjoon and Park, In-Cheol},
  title   = {{Hardware-Efficient Unified Approximation for Implementing Diverse Smooth Activation Functions}},
  journal = {IEEE Transactions on Computers},
  year={2026},
  volume={75},
  number={5},
  pages={2106-2113},
  doi     = {10.1109/TC.2026.3661496}
}

@article{DongPLAC,
  author  = {Dong, Hongxi and Wang, Manzhen and Luo, Yuanyong and Zheng, Muhan and An, Mengyu and Ha, Yajun and Pan, Hongbing},
  title   = {{PLAC: Piecewise Linear Approximation Computation for All Nonlinear Unary Functions}},
  journal = {IEEE Transactions on Very Large Scale Integration (VLSI) Systems},
  year={2020},
  volume={28},
  number={9},
  pages={2014-2027},
  doi     = {10.1109/TVLSI.2020.3004602}
}

@inproceedings{Carrillo2026SoftmaxCORDIC,
  author       = {Carrillo-Arroyo, Andres and Leiva-Valverde, Anthony and Morales-Monge, Roger and Soto-Porras, Carlos and Leon-Vega, Luis G. and Castro-God{\'i}nez, Jorge},
  title        = {{Evaluating CORDIC-Based and Approximate Softmax Accelerators for Deep Learning Inference on FPGAs}},
  year         = {2026},
  booktitle      = {VII Jornadas Costarricenses de Computación e Informática (JoCICI)}
}

@INPROCEEDINGS{BasicSplit,
  author={Sun, Qiwei and Di, Zhixiong and Lv, Zhengyang and Song, Fengli and Xiang, Qianyin and Feng, Quanyuan and Fan, Yibo and Yu, Xulin and Wang, Wenqiang},
  booktitle={2018 14th IEEE International Conference on Solid-State and Integrated Circuit Technology (ICSICT)}, 
  title={A High Speed SoftMax VLSI Architecture Based on Basic-Split}, 
  year={2018},
  volume={},
  number={},
  pages={1-3},
  doi={10.1109/ICSICT.2018.8565706}}

@article{kachris2025survey,
  author  = {Kachris, Christoforos},
  title   = {A Survey on Hardware Accelerators for Large Language Models},
  journal = {Applied Sciences},
  volume  = {15},
  number  = {2},
  pages   = {586},
  year    = {2025},
  doi     = {10.3390/app15020586}
}

@INPROCEEDINGS{leon2022dcas,
  author={León-Vega, Luis G. and Salazar-Villalobos, Eduardo and Castro-Godínez, Jorge},
  booktitle={2022 IEEE 15th Dallas Circuit And System Conference (DCAS)}, 
  title={{An Exploration of Accuracy Configurable Matrix Multiply-Addition Architectures using HLS}}, 
  year={2022},
  volume={},
  number={},
  pages={1-6},
  doi={10.1109/DCAS53974.2022.9845501}
}

@article{leon2023tecs,
author = {Le\'{o}n-Vega, Luis G. and Salazar-Villalobos, Eduardo and Rodriguez-Figueroa, Alejandro and Castro-God\'{\i}nez, Jorge},
title = {{Automatic Generation of Resource and Accuracy Configurable Processing Elements}},
year = {2023},
issue_date = {July 2023},
publisher = {Association for Computing Machinery},
address = {New York, NY, USA},
volume = {22},
number = {4},
issn = {1539-9087},
doi = {10.1145/3594540},
journal = {ACM Trans. Embed. Comput. Syst.},
month = jul,
articleno = {75},
numpages = {27}
}

\end{document}